\documentclass[
 reprint,
 superscriptaddress,
 amsmath,amssymb,
 aps,
]{revtex4-2}

\usepackage{graphicx}
\usepackage{dcolumn}
\usepackage{bm}

\begin{document}

\title{Tracking exciton diffusion across distance and time in rubrene microcrystals
}

\author{Moha Naeimi}
\affiliation{Institute of Physics, University of Rostock,  Germany}
\affiliation{Department Life, Light \& Matter, University of Rostock,  Germany}

\author{Ingo Barke}
\affiliation{Institute of Physics, University of Rostock,  Germany}
\affiliation{Department Life, Light \& Matter, University of Rostock,  Germany}

\author{Stefan Lochbrunner}
\affiliation{Institute of Physics, University of Rostock,  Germany}
\affiliation{Department Life, Light \& Matter, University of Rostock,  Germany}

\author{Sylvia Speller}
\email{Corresponding author: sylvia.speller@uni-rostock.de}
\affiliation{Institute of Physics, University of Rostock,  Germany}
\affiliation{Department Life, Light \& Matter, University of Rostock,  Germany}

\begin{abstract}
Exciton diffusion in organic semiconductors strongly depends on molecular packing and excitation conditions. Here, we investigate spatial exciton diffusion and diffusion dynamics in orthorhombic rubrene crystals. We determine the diffusion coefficients in the geminate time regime, in which the recombination of spin-correlated triplet-pair states dominates, and in the non-geminate recombination time regime, upon random-walk of independently diffusing triplet excitons. We show that the geminate recombination is associated with the lifetime of 3.5 ns and the dominant emission band centred at 650 nm. Moreover, the diffusion in the non-geminate regime is strongly anisotropic, with diffusion coefficients differing by orders of magnitude, depending on the crystallographic direction. Finally, the 650 nm band is proposed as a signature for geminate recombination of the triplet-pair state. Diffusion is also supportive for the observation of quantum beating, i.e., periodic oscillations in slightly delayed fluorescence, caused by superposition of spin-coherent triplet pairs from singlet fission.

\end{abstract}

\maketitle

\section{Introduction}

Organic semiconducting molecular crystals have been proven to be a promising platform for efficient charge and energy transfer via photoexcitation, due to their flexible non-covalent intermolecular electronic interactions and wavefunction overlaps \cite{Wang2018, Mller2023, Najafov2006}. Exciton diffusion in organic crystals is strongly governed by molecular packing, structural order, and the interplay between singlet fission and triplet dynamics. Rubrene (H$_{42}$C$_{28}$) as a semiconductor crystal is known to have not only one of the highest charge mobilities among organic molecules in crystalline form, but also exhibits interesting anisotropic charge and exciton diffusion properties \cite{Stehr2011,biaggio2026review}. A striking feature of rubrene crystals is the efficient singlet fission \cite{Felter2019}, an ultrafast, spin allowed, non-radiative process in which an excited singlet exciton converts into two lower energy triplet excitons. This is energetically favourable when the energy of the first electronically excited singlet state of the molecule ($S_1$) measures about two times the energy of the first electronically excited triplet state ($T_1$). Rubrene crystals exhibit near resonant energetics for singlet fission i.e., the energy of S$_1$ state is approximately twice that of the triplet state T$_1$ \cite{Baronas2022, Naeimi2026, Ma2012}. This energetically resonant process results in the conversion of a singlet exciton into a correlated triplet-pair state (TT), eventually undergoing geminate (TT) or non-geminate (T...T) recombination. The population of the triplet-pair state (since it emerges from a singlet) exhibits a strong dependency on the excitation polarisation, with the highest when the excitation polarisation is aligned with the transition dipole moment \cite{Naeimi2025} -along the longest crystallographic axis- usually referred to as \textbf{c}-axis. On the other hand, singlet fission occurs on picosecond timescales, with the triplet generation dynamics depending on the excitation energy \cite{Finton2019}, while time-resolved photoluminescence measurements reveal delayed emission on much longer timescales associated with triplet–triplet annihilation \cite{Wen2013}.

Spatially resolved measurements have shown highly anisotropic exciton diffusion over micrometer length scales, with triplet diffusion lengths of $4.0\pm0.4\mu$m, attributed to long range triplet diffusion \cite{Irkhin2011} along the shortest crystallographic axis- here referred to as \textbf{a}-axis. The magnitude and range of exciton diffusion are highly dependent on structural disorder and exciton lifetime. In rubrene single crystals, defects substantially reduce the triplet lifetime and corresponding triplet diffusion length\cite{Irkhin2016}. For disordered rubrene films, a singlet diffusion coefficient of $\sim$2.3$\times10^{-3}$cm${^2}$s$^{-1}$ is reported, while the triplet diffusion coefficient measures only $\sim$3.4$\times10^{-7}$cm${^2}$s$^{-1}$, with triplet diffusion lengths of only tens of nanometres \cite{Radiunas2022, Tamura2019}. For crystalline orthorhombic rubrene, diffusion coefficients of 3.5$\times 10^{-3}$cm$^2$s$^{-1}$ and 2.3 cm$^2$s$^{-1}$ are measured for triplet and singlet excitons, respectively, along the \textbf{a}-axis \cite{zhu2018huang}.

The photoluminescence bands in orthorhombic rubrene have been intensively studied with respect to crystal orientation \cite{Irkhin2011, Tavazzi2007, Mitrofanov2007, Naeimi2025}, vibronic progression \cite{Huang2010}, structural inhomogeneity \cite{Chen2011, Bossanyi2024} and singlet fission dynamics \cite{Ma2013}. The high-energy emission around 560-570 nm is associated with an intrinsic, \textbf{c}-polarized electronic transition of crystalline rubrene \cite{Irkhin2012}. The neighbouring 600–610 nm band has been identified as this transition involving a 0-1 vibronic excitation \cite{Huang2010}. Polarisation resolved spectroscopy shows that the relative intensities of these bands depend strongly on the crystallographic orientation \cite{Naeimi2025} and excitation geometry \cite{Irkhin2012}.

Moreover, low-energy emission bands centred near 650 nm and 700 nm were repeatedly reported\cite{Ma2013, Ma2012, Naeimi2025, Park2011, Chen2011}. They are generally not considered as intrinsic molecular emission, but rather are attributed to emissive species that are governed by either structural inhomogeneities \cite{Chen2011, Bossanyi2024} or oxidation \cite{Mitrofanov2006, Mitrofanov2007, Kloc2009} perturbing the structures of the crystal lattice. Ultrafast spectroscopy indicates that the 650 nm and 700 nm emissions exhibit a delayed rise \cite{Baronas2022, Ma2013, Bai2014} in their intensity that kinetically tracks the decay of the higher energy singlet emission. This temporal evolution strongly suggests an energy transfer process, where population is fed from the intrinsic excitonic into lower energy states \cite{Baronas2022}. Open questions concern the nature of the 650 nm peak and its role in geminate triplet pair fusion, exciton diffusion, non-local luminescence, and conditions enabling the observation of quantum beating.

 This motivates the direct investigation of the spatiotemporal evolution and crystallographic anisotropy of exciton diffusion in well-defined rubrene single crystal morphologies, where the relationship between optical excitation, exciton dynamics, and long range diffusion can be resolved simultaneously. As discussed in our previous work \cite{Naeimi2025}, the growth habit of rubrene crystals along the \textbf{b}-axis (\textbf{b}-oriented tabular crystals) can result in the formation of zone-sectored crystals. The zone-sectoring originates from differences in the unit cell orientation around the crystallographic \textbf{a}-axis, leading to different out-of-plane components of the \textbf{c}-axis. These crystallographic differences give rise to distinct exciton dynamics and consequently photoluminescence spectra. Specifically, two characteristic sectors are observed: one with the \textbf{c}-axis dominantly within the crystal plane, and another with the \textbf{c}-axis partially tilted out of the plane. The former exhibits a dominant, strongly polarisation dependent photoluminescence band centred at 650 nm ~\cite{Naeimi2025}. Though this signature was usually differently attributed we hypothesised that it is associated with spin-correlated triplets in geminate pairs. It is strongly \textbf{c}-polarized and characterized by a mono-exponential decay of approximately 3.8 ns. In contrast, the other sector is characterized by dominant, ultrafast decaying emission bands centred at 564 nm and 600 nm and a weak band at 650 nm. These different crystallographic zones with their distinct spectral signatures allow for the study of how the local excited state population influences the subsequent exciton diffusion. In the present work, we investigate the exciton diffusion in rubrene crystals by real-space observations, employing an spatially decoupled excitation-emission measurement scheme. The key result of this work is that the exciton diffusion is highly anisotropic with the crystallographic \textbf{a}-axis as the axis toward which, the exciton transfer is faster and reaching further.

\section{Experimental section}

Sample preparation is explained in details in our previous paper \cite{Naeimi2025}. Spatiotemporal imaging was performed in a confocal microscope setup (MicroTime 200, PicoQuant) using with a time correlated single photon detector, were the excitation and detection are spatially decoupled via a galvo scanner. The excitation sources where 440 nm and 532 nm picosecond pulse lasers operating at 250 kHz directly applied to the sample by a 20x objective lens. The emission was collected by the same objective passed through a galvo scanner to make sure that the detection is scanned over the sample area. The setup can be configured for spatially coupled excitation–detection by directing the excitation light through the galvo scanner using a flipping mirror.

\section{Results and discussion}

\begin{figure}
    \centering
    \includegraphics[width=1\linewidth]{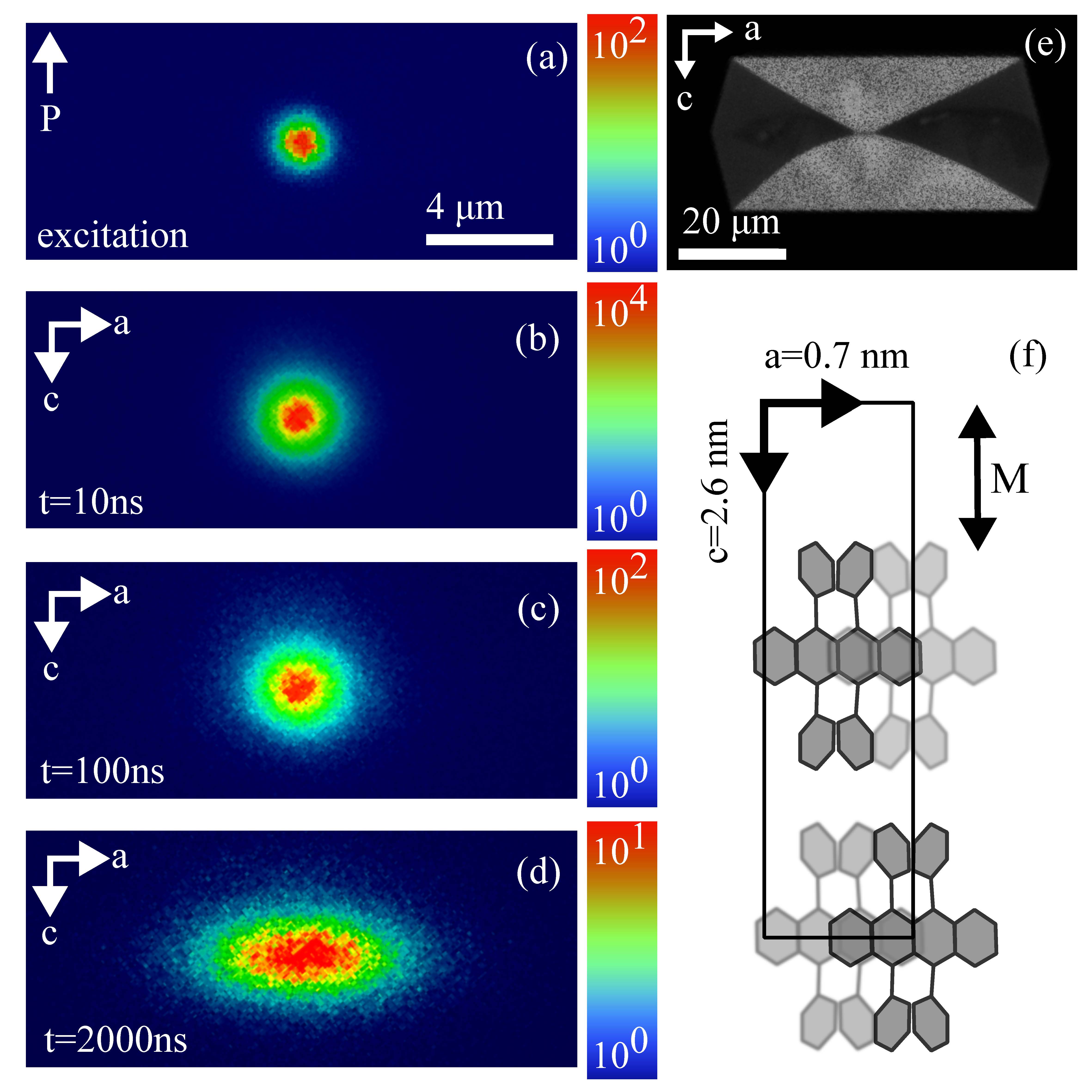}
    \caption{Exciton diffusion in a rubrene crystal. (a) Intensity profile of the excitation light with vertical polarization direction (indicated by arrow), measured on a pristine silicon substrate. (b-d) Intensity profiles at different times after excitation, visualising exciton diffusion. (e) Fluorescence image of a \textbf{b}-oriented rubrene crystal showing distinct dark and bright regions. (f) Molecular packing of orthorhombic rubrene in the \textbf{ac}-plane. The transition dipole moment is perpendicular to the tetracene backbone along the \textbf{c}-axis.}
    \label{fig:diff}
\end{figure}

In Figure \ref{fig:diff}, the spatio-temporal evolution of the emission of a zone-sectored \cite{Naeimi2025} rubrene crystal (shown in Figure \ref{fig:diff}e) after optical excitation at 532 nm is presented. Figure \ref{fig:diff}a shows the excitation spot which is fixed in the centre of the detection window. The time evolution of the spatial distribution of the emission is depicted in Figures \ref{fig:diff}b to \ref{fig:diff}d, showing the expansion of the exciton distribution from the excitation centre. With increasing delay time, the distribution gradually broadens along the \textbf{a}-axis, while only a negligible expansion is observed along the \textbf{c}-axis. This anisotropic evolution directly reflects the highly anisotropic nature of exciton diffusion in rubrene crystals \cite{Irkhin2011}. The molecular packing of the rubrene molecules in the \textbf{ac}-plane of the orthorhombic crystal is illustrated in Figure \ref{fig:diff}e. The transition dipole moment of the rubrene molecules is along the short axis of the tetracene backbone, i.e. along the \textbf{c}-axis. The polarization of the optical excitation is set parallel to the \textbf{c}-axis (Figure \ref{fig:diff}a). The exciton diffusion along the \textbf{a}-axis extends over longer distances, indicating that the diffusion is strongest along the direction with denser molecular packing.

\begin{figure}
    \centering
    \includegraphics[width=1\linewidth]{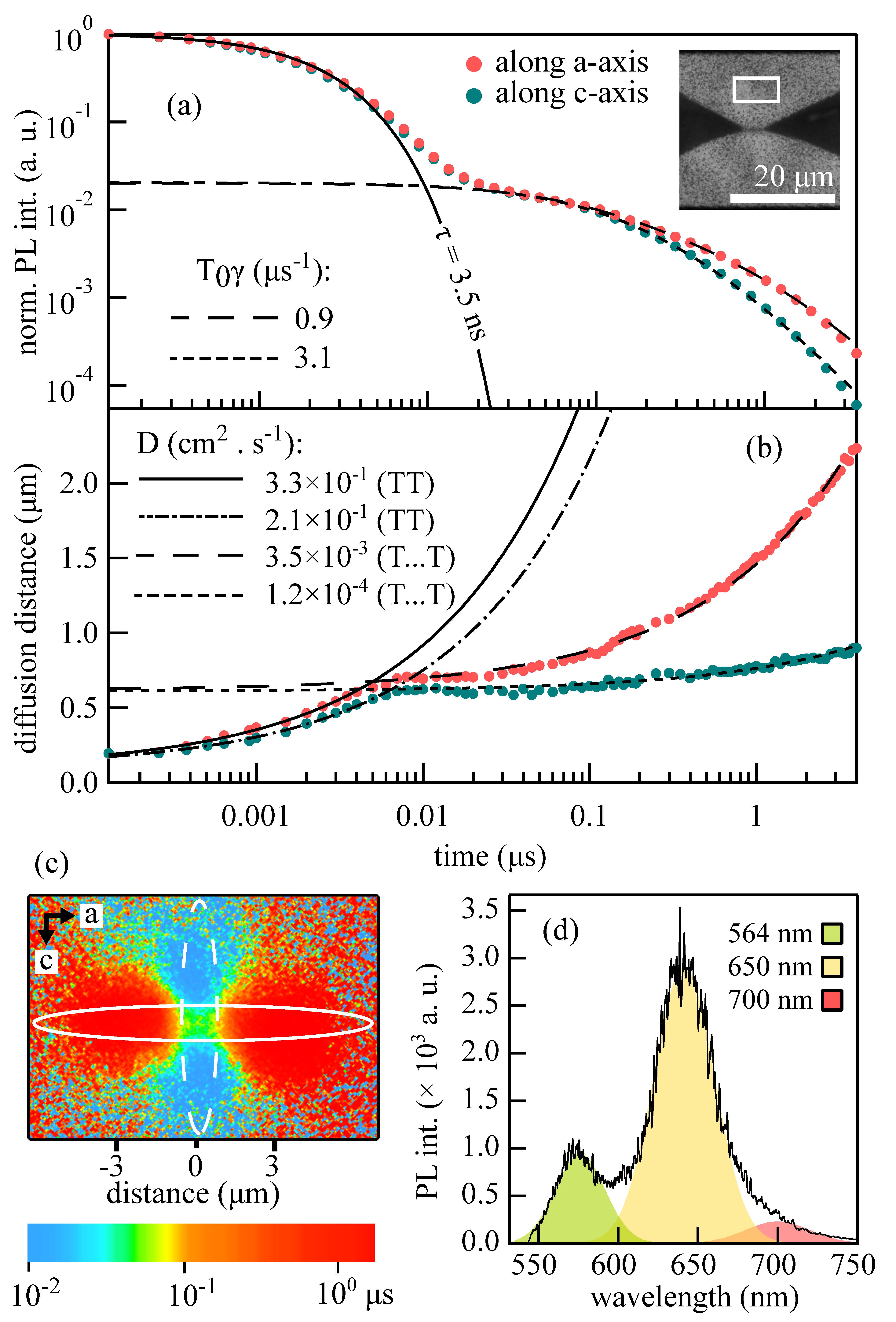}
    \caption{Exciton dynamics and diffusion in rubrene crystals in zones with a dominant 650 nm photoluminesce band (bright zones in Figure \ref{fig:diff}e). (a) Time dependent photoluminescence intensity resolved along \textbf{a}- and \textbf{c}-axis, respectively, corresponding to the red and light-blue regions in panel c. Inset: Microscope image of the rubrene crystal, showing the region from which the data were measured. The dashed curves are fits obtained by Equation \ref{EqRW}. The data are smoothed with first-order locally estimated scatterplot smoothing (LOESS) and subsequently binned for quasi distant point spacing on a logarithmic scale for improved visualization; the raw data are available in the data repository. (b) Excitation-corrected standard deviation of photoluminescence intensity versus time after excitation. The solid and dashed black curves are fits obtained by Equation \ref{EqDiffcoef}. (c) The corresponding lifetime map of photoluminescence distribution. The region corresponding to the lifetime map is indicated in the inset of panel a. (d) Photoluminescence spectrum of zones with dominant 650 nm band.}
    \label{fig:T}
\end{figure}

The spatial broadening of the emission intensity distribution is analysed using the excitation-corrected standard deviation:

\begin{equation}
\sigma_j=\sqrt{
\frac{\sum_j I_j(x_j-x_0)^2}
{\sum_j I_j} - \sigma^2_{ex}}
\label{EqSMD}
\end{equation}

where $I_j$ is the emission intensity at the spatial coordinate ($x_j$). $x_0$ represents the centre of the excitation and $\sigma_{exc}$ the width of the excitation profile (Figure \ref{fig:diff}a). Unlike Gaussian fitting, the excitation-corrected standard deviation is a model independent analysis of the distribution and remains valid even when the spatial profile deviates from a Gaussian shape during the diffusion process.

Using this evaluation approach, the diffusion distance versus delay time can be addressed. Figure \ref{fig:T}a and \ref{fig:T}b show the photoluminescence kinetics and diffusion dynamics of zones (bright regions in Figure \ref{fig:diff}e) where the emission spectra are dominated by the band at 650 nm (as shown in Figure \ref{fig:T}d) plotted along the \textbf{a}- and \textbf{c}-axis (corresponding to solid and dashed marked regions in Figure \ref{fig:T}c, respectively). The time evolution of the exciton dynamics and diffusion exhibits two distinct time regimes. During the first $\sim$ 10 nanoseconds after excitation, the photoluminescence kinetics are described by a mono-exponential decay with a characteristic lifetime of $\sim$ 3.5 ns; this holds for both directions with very similar time constants. Interestingly, the diffusion distance $L=\sigma$ reaches during this time interval of $\sim$ 10 ns a magnitude of  $\sim$ 0.5 $\mu$m simultaneously along both axes, as shown in Figure \ref{fig:T}b. The diffusion distance is the root mean square of the time-dependent spatial distribution of the diffusion equation ($\frac{\partial^2 X}{\partial^2 x}=D\frac{\partial X}{\partial t}$) with $X$ denoting the population of the excited species under observation and $D$ is diffusion coefficient \cite{Mikhnenko2015}:

\begin{equation}
L(t)=\sqrt{2Dt},
\label{EqDiffcoef}
\end{equation}

Fitting this equation to the diffusion for the first 10 ns, the early time diffusion coefficient $D_{TT}$ is obtained as $\sim$ (2.1 $\pm$ 0.1) $\times 10^{-1}$cm$^2$s$^{-1}$ along \textbf{a}-axis and $\sim$ (3.3 $\pm$ 0.08) $\times 10^{-1}$cm$^2$s$^{-1}$ along \textbf{c}-axis. The interrelation between the characteristic decay time and the spatial diffusion distance suggests that the initial population (till 10 ns) exhibits a high diffusion constant. This is in line with typical singlet diffusion constants. This is during the geminate triplet recombination phase, therefore we refer to it as geminate triplet pair with singlet character. This entangled triplet pair behaves like a singlet species, which can diffuse via hops mediated by a mechanism similar to Förster energy transfer. Although singlet exciton diffusion has been measured in disordered rubrene films, with reported diffusion coefficients of approximately 2.3$\times10^{-3}$ cm$^2$s$^{-1}$ \cite{Radiunas2022} and 2.3 cm$^2$s$^{-1}$ in single crystals \cite{zhu2018huang}, experimental data for singlet diffusion in rubrene crystals are still limited. Theoretical studies have suggested that singlet–triplet interconversion, including singlet regeneration through triplet–triplet annihilation, can contribute to long-range exciton transport in rubrene crystals \cite{Tamura2019}.

For delay times after 100 ns the exciton dynamics becomes very anisotropic (Figure \ref{fig:T}a) and the kinetics are well described by the random-walk recombination model \cite{Ryasnyanskiy2011}:

\begin{equation}
I(t)\propto
\left(1+T_0\gamma t\right)^{-2},
\label{EqRW}
\end{equation}

where $\gamma$ is the non-geminate triplet recombination rate and $T_0$ the initial triplet population. By fitting using this model, values for $T_0\gamma$ of (0.90 $\pm$ 0.01) $\mu$s$^{-1}$ are obtained for various samples along the \textbf{a}-axis and of (3.10 $\pm$ 0.05) $\mu$s$^{-1}$ along the \textbf{c}-axis. Although the initial triplet exciton population T$_0$ and recombination rate $\gamma$ are the same along both directions, the different values obtained for T$_0\gamma$ at longer (t$>0.1 \mu$s) are likely due to the directional dependent annihilation processes, in accordance to the different exciton densities observed in Figure \ref{fig:T}a.

In principle, using data acquired in the spatially decoupled excitation-emission scheme, the extraction of the recombination rates could become corrupted by unbalanced inflow and outflow of triplet excitons into the evaluated region of interest. Therefore, equation \ref{EqRW} may be under-complex for this situation. To assess a possible inaccuracy, we compared the outcome integrated over regions of interest only covering the longer lifetime lobes of the luminescence maps with one large, axially extended region (Figure \ref{fig:T}c). The extracted fitting constants appear largely robust; therefore, we did not apply a dedicated adaptation of the differential equations introduced before \cite{Naeimi2025} for the analysis of the data acquired in the coupled excitation emission scheme.
 
A corresponding trend is observed in the diffusion distance. Along the \textbf{c}-axis, the diffusion is almost negligible after the diffusion in the initial phase. On the other hand, the diffusion distance along the \textbf{a}-axis continues to increase and is well described by the Fickian diffusion model introduced above, yielding diffusion coefficients in the range of
D=(3.5 $\pm$ 0.01)$\times10^{-3}$cm$^2$s$^{-1}$ along the \textbf{a}-axis. For the \textbf{c}-axis, however, a diffusion coefficient in the order of (1.2 $\pm$ 0.005)$\times$10$^{-4}$ cm$^2$s$^{-1}$ is obtained, which is almost one order of magnitude smaller compared to that along the \textbf{a}-axis (see Figure \ref{fig:T}b).

\begin{figure}
    \centering
    \includegraphics[width=1\linewidth]{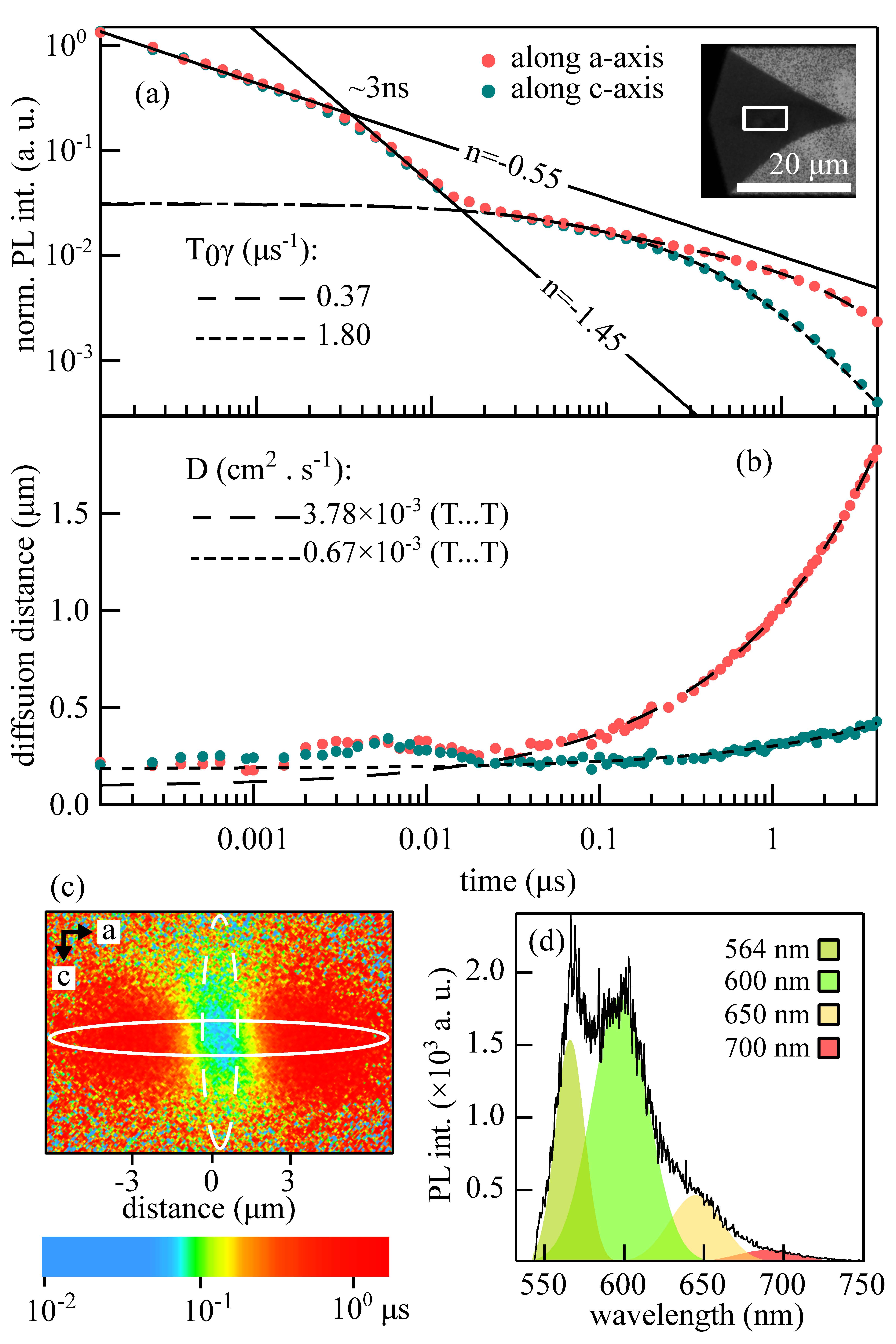}
    \caption{Exciton dynamics and transport in rubrene crystals in zones with a dominant 564 nm bands (dark zones in Figure \ref{fig:diff}e). (a) Photoluminescence intensity versus time after excitation along \textbf{a}- and \textbf{c}-axis, corresponding to the red and green regions in Figure c, respectively. Inset: Microscope image of the rubrene crystal, showing the region from which the data were measured. The dashed curves are fits obtained by Equation \ref{EqRW}. The data are smoothed with LOESS and subsequently binned for quasi distant point spacing on a logarithmic scale for improved visualization; the raw data are available in the data repository.(b) Excitation-corrected standard deviation of photoluminescence intensity versus time after excitation. The dashed black curves are fits obtained by Equation \ref{EqDiffcoef}. (c) The corresponding lifetime map of photoluminescence distribution. The region corresponding to the lifetime map is indicated in the inset of panel a.(d) Photoluminescence spectra of zones with dominant 565 nm and 600 nm bands.}
    \label{D}
\end{figure}

A triplet diffusion constant of 3.5 $\times10^{-3}$ cm$^2$s$^{-1}$ is reported for the fast axis \cite{Zhu2016, zhu2018huang} of rubrene, which matches the value extracted in the present work. Alternatively, triplet exciton hopping times of 3 ps, 250 ps, and 2.6 $\mu$s along the three axes in rubrene, as reported in \cite{Wolf2021, biaggio2026review}, can be transformed to diffusion constants using equation \ref{EqDiffcoef} and taking $L$ as the length of the rubrene orthorhombic lattice constants. This yields a somewhat lower value of D$_a$=8.6 $\times10^{-4}$ cm$^2$s$^{-1}$ compared to the present work. For the \textbf{b}- and \textbf{c}-directions such transformation yields D$_b$=4.1 $\times10^{-5}$ cm$^2$s$^{-1}$ and D$_c$=$1.4\times10^{-8}$ cm$^2$s$^{-1}$ is estimated. With this crystal orientation, the diffusion constant for the \textbf{b}-axis cannot be extracted because it is out of plane. For the \textbf{c}-axis, the diffusion of triplet excitons is generally low, due to the phenyl groups lowering the orbital overlap of neighbour molecules, and thereby Dexter transport. In the present work it is 30 times lower than along \textbf{a}-axis and considerable four orders of magnitude larger compared to the estimated value D$_c$ from reported hopping time. This might be related that the linear thermal expansion coefficient is about four times larger along \textbf{c}, i.e., very little temperature differences due to laser illumination will considerably alter the orbital overlap.

Tetracene is less anisotropic and diffusion constants of 0.17 cm$^2$s$^{-1}$ for singlet excitons and 3.2$\times 10^{-4}$ cm$^2$s$^{-1}$ for triplet excitons undergoing non-geminate recombination, have been calculated \cite{Yost2012}. However, experimental studies reported substantially higher singlet (prior to singlet fission) diffusion constants for tetracene in the range of 20-50 cm$^2$s$^{-1}$ \cite{Muth2024} and 3.1 cm$^2$s$^{-1}$ \cite{Zhu2016, zhu2018huang}. Therefore, the extracted singlet diffusion constant of rubrene is lower than that of tetracene while the triplet diffusion constant is one order of magnitude higher.

The corresponding measurements performed in zones dominated by the 564 nm and 600 nm emission (dark regions in Figure \ref{fig:diff}e) are summarized Figure \ref{D}. In contrast to the regions with dominant 650 nm emission band, no initial mono-exponential decay is observed, due to the ultrafast decay of the initially excited singlet state \cite{Naeimi2025, Ma2013}. Instead, in this regime, the early time dynamics exhibit two consecutive power-law regimes. For times shorter than $\sim$ 3 ns, the photoluminescence intensity follows $I(t)\propto t^{-0.5}$ and then till $\sim$ 20 ns, it follows $I(t)\propto t^{-1.5}$. A striking feature with the two distinct power-law regimes is, that the two exponents do add to -2, and the crossover time is close to the “lifetime” of 3.5 ns. This is indicative of continuous time random walk based on Montroll-Weiss \cite{Montroll1965, Shlesinger2017}. In this framework, the sum of -2 is an universal constant, resulting from the joint sub-diffusion parameter $\beta$ connecting the pre-transit and post-transit power-law exponents $\beta-1$ and $-\beta – 1$. Thus, this points to an anomalous diffusion with a strongly sub-diffusive $\beta$ exponent of 0.45. This implies that ordinary Fickian diffusion cannot evolve. Possibly, a certain degree of energetic disorder or memory is preserved and waiting times between jumps can become long.

In this sector with dominant 564 nm and 600 nm emission the singlet-like diffusion is almost fully suppressed (see Figure \ref{D}b).
It could mean physical or energetic obstacles, such as trapping, may hinder free hopping of geminate triplet pairs, i.e., of singlet-like excitons. This power-law behaviour is exclusively observed when the wavelength of the emitted photons stays short (600 nm and 564 nm), i.e., in the case of low vibrational quantum numbers (0-0 and 0-1).

In contrast, mono-exponential decay indicates a memoryless, Markovian process and singlet-like diffusion is appreciable in both lattice directions, also along the \textbf{c}-axis (Figure \ref{fig:T}). Note, Förster transfer is mediated electromagnetic and does not require electronic overlap between wave-functions of neighbouring molecules. This may partially explain why hardly an anisotropy is observed due to the bulky phenyl groups.  While the Montroll-Weiss model originally has been applied to charge carrier transport in disordered systems \cite{Hughes1996, Pfister1978, Shlesinger2017}, it is applicable to transport of other excitations such as excitons \cite{blumen2011}. The close agreement of the measured exponents suggests that a similar interaction with the environment than with charge carriers corrupts the transport. An option is transient trapping.

At later time regimes, the kinetics observed for zones with dominant intrinsic (i. e. 564 nm and 600 nm ) emission bands turn into the same random-walk recombination dynamics as the zones with dominant geminate triplet-pair emission. Also the obtained fitting values for T$_0\gamma$ are substantially smaller, i.e. (0.37 $\pm$ 0.01) $\mu$s$^{-1}$ for \textbf{a}-axis and (1.80 $\pm$ 0.08) $\mu$s$^{-1}$ for \textbf{c}-axis. This is consistent with our previously reported values \cite{Naeimi2025}, indicating that the parameter T$_0 \gamma$ is smaller in the zone with dominant 564 nm and 600 nm bands, compared to zones with 650 nm bands. The times $1/T_0\gamma$ demarcate the time regime with a sum raised to a power of -2 and the that one when the dynamics takes on a power law with exponent -2 \cite{Ryasnyanskiy2011}. Simultaneously, the triplet exciton diffusion becomes identical to the zones with 650 nm bands: diffusion remains extremely weak along the \textbf{c}-axis with a diffusion constant of D=(0.67 $\pm$ 0.002)$\times10^{-3}$cm$^2$s$^{-1}$, whereas along the \textbf{a}-axis strong diffusion with a diffusion coefficient of about D=(3.78 $\pm$ 0.003)$\times10^{-3}$cm$^2$s$^{-1}$ is found, comparable to those obtained for the 650 nm dominated zones. Both types of zones experience identical long-time triplet exciton diffusion suggesting that despite their distinct initial dynamics, triplet exciton diffusion results in the same population of mobile triplet excitons responsible for long range diffusion.

The results of the present work reveal an intriguing link between three features occurring exclusively together: dominance of the redshifted 650 nm band, mono-exponential decay with lifetime $\sim$ 3.5 ns, and high diffusion constants for triplet-pairs with singlet character. Having in mind, a lifetime of $\sim$ 4 ns can be regarded as a signature of the multiexciton triplet-pair state \cite{biaggio2018, biaggio2026review}, this strongly suggest that geminate triplet-pair recombination leads to emission of the 650 nm luminescence band. Till their recombination the triplet-pair and its diffusion are singlet-like. This is consistent with the observation of almost isotropic diffusion along both the crystallographic \textbf{a}- and \textbf{c}-axis, with a remarkably high diffusion constant of $\sim 2.6 \times 10^{-1}$cm$^2$s$^{-1}$ in sectors with dominant 650 nm luminescence and a mono-exponential lifetime of $\tau_{TT}=$ 3.5 ns. A diffusion length $L_{TT}=\sqrt{D\tau}$ of $\approx$ 118 nm is obtained. The observed mono-exponential decay of the highly mobile singlet-like species with a characteristic lifetime of 3.5 ns indicates the successive recombination of two geminate triplet excitons, leading to the decay with 650 nm emission. In this early time regime spin-correlated triplet pairs with singlet character and possibly ordinary singlets co-exist. Obviously, the diffusion is hardly hampered compared to the other sector when singlet-like diffusion is largely suppressed. Therefore trapping is rather not an explanation for the observed redshift. Currently, it remains obscure why the redshifted signature is associated with decent transport, and the emission with low shift with hampered transport.

Transport processes in rubrene are governed by the dense molecular packing along the \textbf{a}-axis. In that respect, the quasi isotropic, high-mobility dynamics observed for singlet-like excitons appears surprising. It contrasts with the strongly anisotropic transport typically found at later times with non-geminate triplet exciton recombination. Instead, the observed characteristics are consistent with the dynamics of singlet-like triplet-pair states. Geminate triplet-pair recombination in rubrene has been shown to produce delayed fluorescence with lifetimes of few nanoseconds, depending on the coupling strength and crystal orientation \cite{Zhu2016, Irkhin2011, Irkhin2016}. On the other hand, at larger times, the extracted triplet diffusion constant of the order of $10^{-3}$cm$^{2}$s$^{-1}$ is consistent with the large observed diffusion distances of non-geminate triplet excitons in rubrene due to their long lifetime resulting from reduced recombination probability, compared to the correlated triplet pair \cite{Finton2019, Radiunas2022}. 

\begin{figure}
    \centering
    \includegraphics[width=1\linewidth]{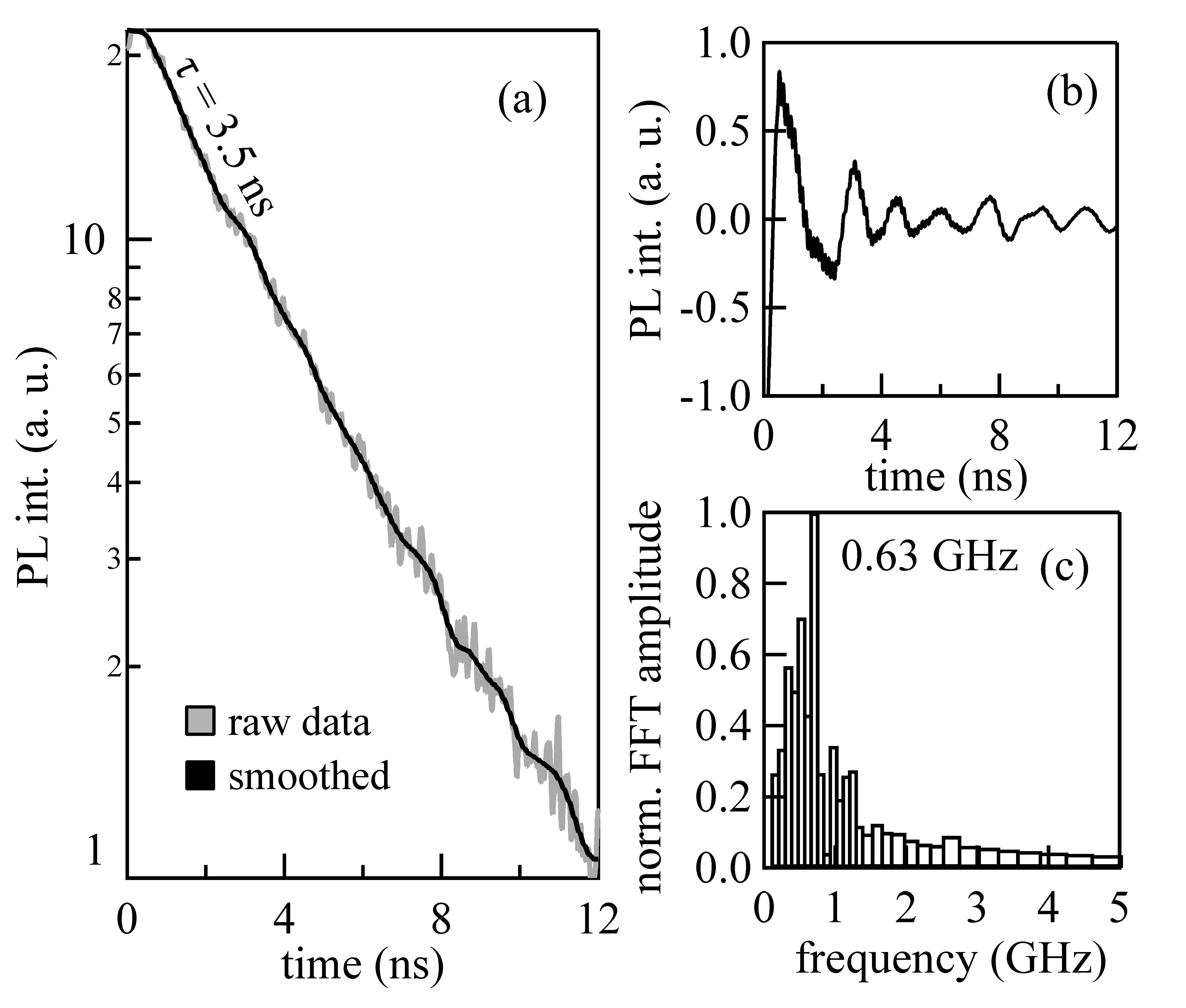}
    \caption{Quantum beats observed on a rubrene crystal, in zones with a dominant 650 nm luminescence band (bright zones in Figure \ref{fig:diff}e), in the absence of an external magnetic field. (a) Photoluminescence intensity versus time after excitation, with exponential time constant of $\sim$3.5 ns. The excitation polarization direction is between \textbf{a}- and \textbf{c}-axes. (b) Relative beating of photoluminescence intensity, obtained by subtracting the monoexponential fit from the photoluminescence raw data from 0 to 12 ns. (c) Fast Fourier transform amplitude versus beating frequency. The dominant beating frequency of 0.63 GHz is observed.}
    \label{QB}
\end{figure}

Another intriguing feature of the \textbf{b}-oriented rubrene crystals is quantum beating. This concerns the early time regime is due to the superposition of triplet exciton pairs with different total spin orientations with respect to the orthorhombic rubrene lattice. Quantum beats can exist in the time regime till $\sim$ 12 ns before geminate triplet fusion becomes exhausted. Beating frequencies are calculated from the zero-field splitting parameters $E$ and $D$ extractable from magnetic resonance experiments. Predicted fundamental beating frequencies for rubrene are 0.48 GHz without Zeeman splitting and 0.6 GHz with external magnetic field \cite{Curran2024, Curran2025}. In orthorhombic rubrene, quantum beats have been observed with the assistance of external magnetic fields, to preserve some coherence during transport \cite{Curran2024}. Simulations revealed that quantum beating depends on hopping rates, and can in principle be observed also without external magnetic fields \cite{Curran2025}. 

Occasionally, we observe quantum beats, without external magnetic fields. Figure \ref{QB}a shows the time dependent photoluminescence intensity of rubrene crystals at zones with dominant 650 nm band. By fitting a monoexponential function, the time constant of $\sim$ 3.5 ns is obtained, similar to the spectra without quantum beating. Subtracting the time dependent decay and its equivalent monoexponential fit, the fluctuations of photoluminescence intensity (i. e. quantum beating) is extracted, as shown in Figure \ref{QB}b. The quantum beating is observed particularly when the polarisation of the laser is in-between \textbf{a} and \textbf{c}, and when using the spatially decoupled excitation - emission measurement scheme. This means photons from the directly excited area are excluded. Note, the strong vibrational energy entry at the excitation area ($\varnothing\approx$ 1um) might destroy quantum coherence and entanglement between geminate triplet pairs with different spin-orientations. Therefore, it is not surprising, that quantum beats can be better observed a little off the directly illuminated centre, and on \textbf{b}-oriented crystals, in absence of externally applied magnetic fields. This is compatible with Monte-Carlo simulations \cite{Curran2025}. The connection with diffusion is that a minimum of hopping or diffusion is supportive to preserve sufficient coherence to observe a decent quantum beating signal outside the directly illuminated area. 

\section{Conclusion}
Real-space singlet and triplet exciton diffusion in rubrene crystals are investigated, revealing a strong correlation between local emission spectra and observed transport dynamics. Regions dominated by the 650 nm band exhibit excitons with an isotropic and high mobility with a diffusion constant of D$\sim 2.5 \times 10^{-1}$cm$^2$s$^{-1}$ along both \textbf{a}- and \textbf{c}-axis) and with a mono-exponential lifetime of 3.5 ns, attributed to geminate recombination of spin-correlated triplet-pair states. In contrast, regions dominated by the intrinsic 564 and 600 nm emission bands show negligible early transport. At later times, a long range anisotropic triplet diffusion along the \textbf{a}-axis with D$\sim 3 \times 10^{-3}$cm$^2.$s$^{-1}$ occurs. These results introduce the 650 nm band as a fingerprint of geminate triplet-pair recombination and highlight the distinct transport mechanisms of correlated versus single triplet excitons.
Quantum beating in areas nearby but outside laser excitation is observable, suggesting diffusion is key to preserve decent coherence. 

\section{Acknowledgment}
Funding by the Deutsche Forschungsgemeinschaft (DFG, German Research Foundation) within projects SFB 1477 “Light-Matter Interactions at Interfaces” (441234705), SFB 1270 “Electrically Active Implants” (299150580) and “Application of Interoperable Metadata Standards (AIMS) 2” (432233186).

\bibliography{Ref}

\end{document}